\documentclass[twocolumn,
prl,aps,a4paper
,amsfonts,amssymb,bm]{revtex4}
\usepackage[fleqn]
{amsmath}
\usepackage{graphicx}

\begin{document}
\flushbottom 

\title{On phase transitions in $\mathbf{Pb/Ge(111)}$ and $\mathbf{Sn/Ge(111)}$ }

\author{A. Cano}\email{andres.cano@uam.es}
\author{A. P. Levanyuk}\email{levanyuk@uam.es}
\affiliation{Departamento de F\'\i sica de la Materia Condensada, C-III, Universidad Aut\'onoma de Madrid, E-28049 Madrid, Spain}


\begin{abstract}
The present communication is a critical examination of two points relevant to the surface phase transitions of $\rm Pb$ and $\rm Sn$ overlayers on $\rm Ge(111)$. One is connected with the reading of the reported structural data, which lead to some consequences of paramount importance but overlooked up to now. The other point is the seeming contradiction in what concerns to the transition temperatures obtained from different experimental methods. The importance of this contradiction, in contrast to the previous point, has been overestimated. 
\end{abstract}
\pacs{PACS number(s): }
\maketitle


\section{I. Structural data}

Surface phase transitions of $\rm Pb$ and $\rm Sn$ overlayers on $\rm Ge(111)$ have been the subject of numerous studies. 
It is generally assumed that $\sqrt 3 \times \sqrt 3 R 30^{\circ} $ (in the following $\sqrt 3 $) to $3\times 3 $ phase transitions in $\rm Pb/Ge(111)$ and $\rm Sn/Ge(111)$ are completely equivalent (see, e.g., Refs. \cite{Mascaraque99,Bunk99,Melechko99,Floreano01}). However, taking into account the structural data reported in Refs. \cite{Mascaraque99,Bunk99} one has to realize that, in fact, these transitions should be very different. It is because of the following symmetry arguments. On one hand, both $\rm Pb/Ge(111)$ and $\rm Sn/Ge(111)$ systems have $\sqrt 3$ structures with the symmetry of the $p31m$ space group. On the other hand, the symmetry of the $3\times 3$ structure in $\rm Pb/Ge(111)$ is $p3$ \cite{Mascaraque99} while in $\rm Sn/Ge(111)$ it is $p3m1$ \cite{Bunk99}. Therefore, the order parameters of these transitions are different, i.e. they transform according to different irreducible representations of the $p31m$ space group. Moreover, the corresponding Landau-Ginzburg-Wilson (LGW) Hamiltonians, which can be constructed according to Ref. \cite{Rottman81}, are quite different. This leads, in particular, to very different critical behaviors. The LGW Hamiltonian for the $\sqrt3 $ to $3\times 3$ transition in $\rm Sn/Ge(111)$ is that of the three-state Potts model. Experimental data \cite{Floreano01} seem to be in agreement with the critical behavior of this model \cite{Alexander75}. The LGW Hamiltonian for the $\sqrt3 $ to $3\times 3$ transition in $\rm Pb/Ge(111)$ is, however, that of the $XY$ model with sixth-order anisotropy. The critical behavior of this model is completely different \cite{Jose77}. On lowering the temperature one first crosses an upper critical temperature into a Kosterlitz-Thouless region of critical points with continuously variable exponents. One finally crosses a second critical temperature into an ordered phase. 

There exist, however, another simpler possibility. Note that, while the $3 \times 3$ structure in $\rm Pb/Ge(111)$ appears at $\sim 250~\rm K$, the data showing $p3$ symmetry were obtained at $50~\rm K$ \cite{Mascaraque99}. 
Therefore, the total lowering of symmetry could take place in two steps: (i) a first phase transition is responsible of the appearance of a $3\times 3$ structure with $p3m1$ symmetry and (ii) a second phase transition lower this symmetry to $p3$ with no increase of the unit cell. If it were the case, $\sqrt 3$ to $3\times 3$ transitions in both $\rm Pb/Ge(111)$ and $\rm Sn/Ge(111)$ systems would be equivalent. To our knowledge, $\rm Pb/Ge(111)$ has not been studied enough to affirm what of the above described possibilities, if any, takes place. So we underline the necessity of such a study.

\section{II. Transition temperature}

Let us now examine the other point. In Refs. \cite{Melechko99}, the transition temperature of the $\sqrt3$ to $3 \times 3$ transition in $\rm Sn/Ge(111)$ has been established, from STM experiments, as the temperature at which the region of $3\times 3$ structure induced by some defects becomes infinite \cite{nota}. The difference between this temperature and that obtained from diffraction experiments \cite{Floreano01} is remarkably large ($\sim 100^{\circ}$). However if these defects are mobile, what seems to be the case \cite{Melechko99}, the temperature reported in Ref. \cite{Melechko99} has nothing to do with the real transition temperature: the long-range order, what implies defect ordering, is not probed in these experiments. Therefore, the so large difference between these temperatures is not so surprising. Indeed, the influence on the phase transition temperature of mobile symmetry-breaking defects is well known since long ago \cite{Halperin76}. Experimental examples of such a influence, for instance in $\rm KTaO_3:Li$, are also well documented \cite{Vugmeister90}. This crystal is an incipient ferroelectric in absence of $\rm Li$-dopants, i.e. its ``transition temperature'' is negative and close to zero. The $\rm Li$-dopants act as orientable dipoles (mobile symmetry-breaking defects) and, as a result of these new degrees of freedom, the resulting transition temperature is high enough to permit that the phase transition takes place. The corresponding increase of the $\sqrt 3 $ to $3\times 3$ transition temperature in $\rm Sn/Ge(111)$ and $\rm Pb/Ge(111)$ can be conveniently estimated according to Ref. \cite{Levanyuk79}. One obtains that, if the defect concentration is $\sim 1 \%$, this temperature could change up to $\sim 100^{\circ}$ for not necessarily very strong defects.

\subsection{Acknowledgments}
We acknowledge E.G. Michel and A. Mascaraque for useful discussions.


\begin{thebibliography}{999}
\bibitem{nota} Let us mention that the amplitude of the defect-induced modulation is $\sim K_0(\rho/\xi)$, where $K_0$ is the MacDonald function and $\xi$ is the coherence lenght in absence of defects (see, e.g., 
Ref. \cite{Levanyuk79}). 
For $\rho\gg \xi$ this amplitude decays as $(\xi/\rho)^{1/2}\exp (-\rho/\xi)$. In Refs. \cite{Melechko99}, however, a different form $\exp (-\rho/\xi)$ was used to fit the experiments.
\end{thebibliography}
\end{document}